\documentclass[preprint,proceedings]{rmaa}

\usepackage{paralist}

\usepackage{psfrag,color}

\SetYear{2002}
\SetConfTitle{Winds, Bubbles \& Explosions}

\title{Observation and Modelling of Starburst Driven Galactic Winds:
A Review in honour of John Dyson}

\author{
  D. Breitschwerdt\altaffilmark{1}}

\altaffiltext{1}{Max-Planck-Institut f\"ur extraterrestrische
Physik, Garching, Germany.}

\shortauthor{Dieter Breitschwerdt}
\shorttitle{RevMexAA(SC) Starburst Galaxies}

\fulladdresses{
\item Dieter Breitschwerdt: Max-Planck-Institut f\"ur
extraterrestrische Physik, Postfach 1312, D-85741 Garching,
Germany (breitsch@mpe.mpg.de).}

\listofauthors{D. Breitschwerdt}
\indexauthor{Breitschwerdt, D.}

\abstract{Starburst galaxies are generally associated with
extended X-ray and radio halos, giving a clear hint of an outflow
of gas and relativistic particles from the disk into the halo. The
driving agents are, not surprisingly, active star forming regions,
injecting hot gas and cosmic rays generated by supernova remnant
and superbubble shock waves. The dynamics of the outflow and the
thermal evolution of the plasma are strongly coupled, and
therefore a self-consistent model is necessary for a satisfactory
description. Since the plasma temperature is between one and a few
million Kelvin, the halo is most conspicuous in soft X-rays. It
will be shown that X-ray data obtained recently with \emph{XMM-Newton} 
EPIC pn bear clear spectral
signatures that are in strong disagreement with an isothermal halo
in collisional ionization equilibrium (CIE). Instead a temperature
structure in X-ray halos is observed. It will be demonstrated that
a galactic wind outflow model in which the non-equilibrium
ionization structure is calculated self-consistently can give a
satisfactory and physical explanation for the X-ray spectral
characteristics, e.g.\ in case of the local starburst galaxy NGC 3079. 
In particular, high subsolar abundances, that have
been reported from X-ray observations in some cases, are shown to
be artefacts of CIE spectral fit models. It is found that
spectral models are strongly constrained by the presence of
diagnostic lines such as \ion{O}{7} and \ion{O}{8} as well as Fe-L
line complexes. Thus a detailed spectral X-ray model will help to provide 
important galactic wind parameters, such as mass loss rate and velocity 
profile, which may also serve as valuable input for galactic outflows in the 
early universe.}

\addkeyword{hydrodynamics}
\addkeyword{galaxies: halos}
\addkeyword{galaxies: starburst}
\addkeyword{X-rays: galaxies}

\begin{document}
\maketitle

\section{Introduction}
\label{sec:intro} The observation of optically selected late-type
galaxies led to the conclusion that the bluest objects could not
all be young, but more likely are in a state of enhanced star
formation (Searle et al. 1973). These so-called starbursts are
intermittent during galaxy evolution with a duration of typically
$10^8$ yrs. It has been argued that the total rate of high-mass
star formation within these short (with respect to a Hubble time)
periods is comparable in the local universe to the rate in spirals
during the whole time of quiescent star formation (Heckman 1997).
Undoubtedly, the effects of a starburst on the ISM are dramatic. A
high supernova rate disturbs and eventually disrupts the gaseous
disk and supernova remnant (SNR) heated gas and accelerated
energetic particles (cosmic rays) are injected into the base of
the galactic halo.

It has been shown theoretically, that the combined pressure
gradients of gas, cosmic rays (CRs) and MHD waves are able to
drive a galactic wind even in cases of moderate star formation,
like in our own Galaxy (Breitschwerdt et al. 1991). A necessary
condition is that there is a dynamical coupling between the plasma
and the CRs. Such a coupling can be provided by strong resonant
scattering off self-excited MHD waves, which are generated by a
large scale CR gradient pointing away from the galaxy and thus
induce a so-called streaming instability (e.g.\ Kulsrud \& Pearce
1969). On the other hand, the overpressure in superbubbles
generated by rich OB clusters is sufficient to drive a largely
thermal outflow. In case of starburst galaxies the conditions are
even more extreme, and in galaxies like M82, the galactic wind (or
superwind, as it is sometimes called) is purely thermally driven
(Chevalier \& Clegg, 1985; V\"olk et al. 1996).

\begin{figure}[!t]
  \includegraphics[width=\columnwidth]{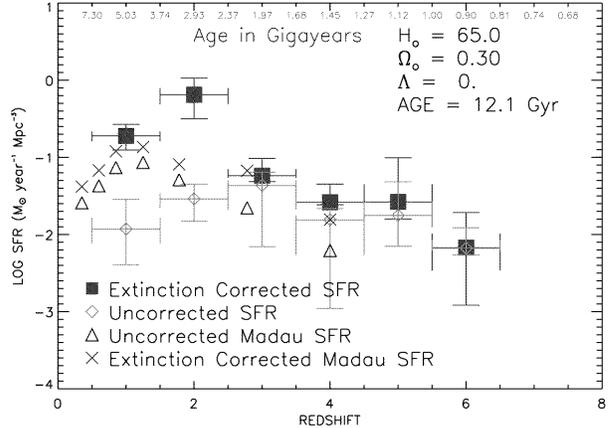}
  \caption{Star formation rate as a function of redshift in a comoving volume
  including also extinction (Thompson et al. 2001) }
  \label{fig:sfr}
\end{figure}

\begin{figure}[!t]
  \includegraphics[width=\columnwidth]{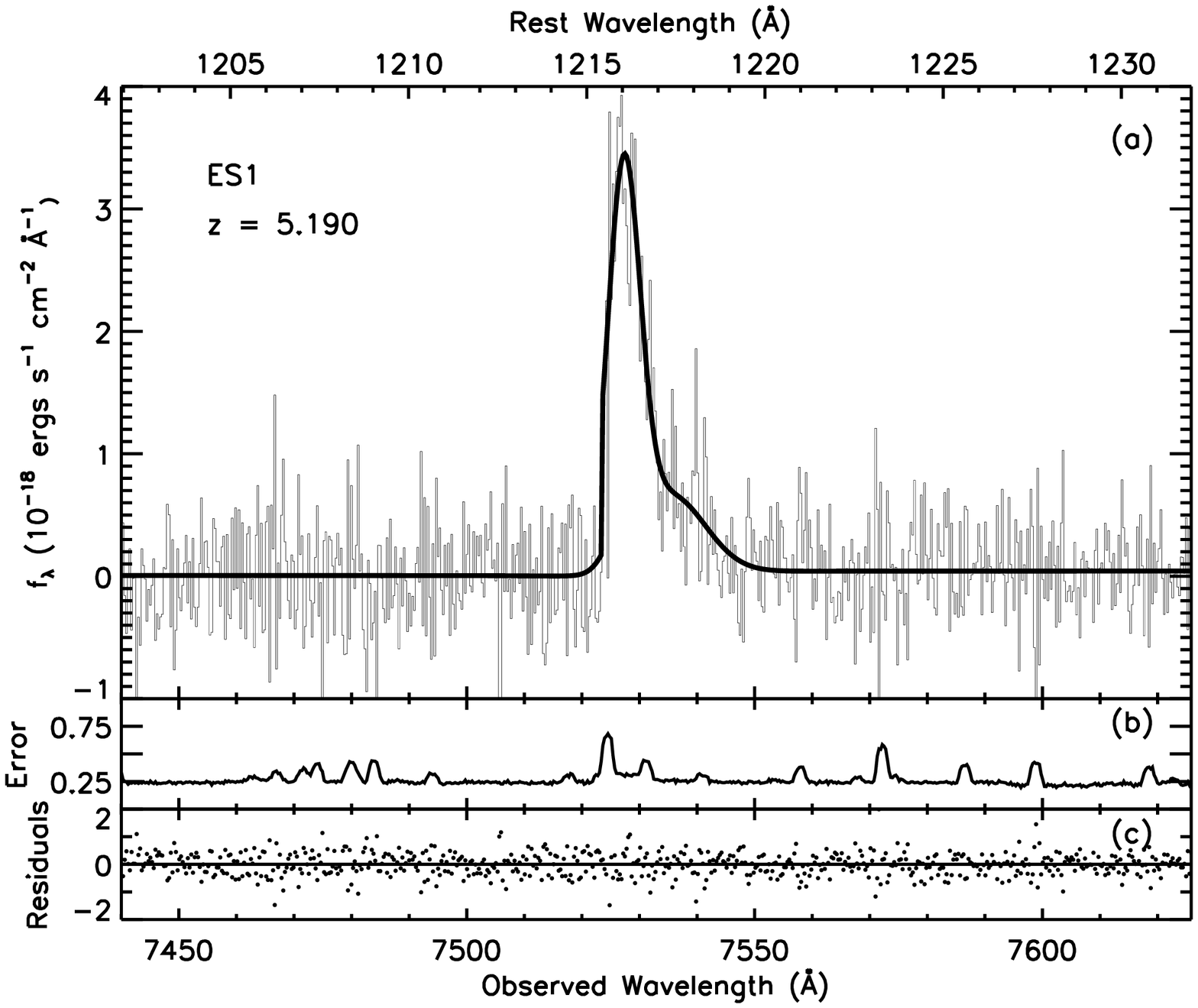}
  \caption{Ly$\alpha$ emission line from the outflow of the star-forming
galaxy J123649.2+621539 (Dawson et al. 2002). The fitted line
profile could be modeled by 3 components: a large amplitude narrow
Gaussian for the recombination of ${\rm H}^+$, a small amplitude
broad Gaussian (redshifted by 320 km/s) for backscattered red-wing
emission of far-side Ly$\alpha$ photons of the expanding wind, and
a broad Voigt absorption profile (blueshifted by 360 km/s) to
model the blue decrement due to near-side absorption of photons. }
  \label{fig:gw_la}
\end{figure}

In recent years the study of starburst phenomena and galactic
winds has received a tremendous boost. This was largely driven by
observations, notably the availability of 8~m-class telescopes such
as the VLT and Keck with superb instrumentation, as well as the
\emph{HST}. Selecting galaxy samples by colour and measurement of
the UV flux of galaxies indicated a peak in the star formation
rate (SFR) near redshift $z \sim 1.5$ (Madau et al. 1996), while in
more recent observations it was argued for a near constancy
between $z=1.5 - 4$ (Steidel et al. 1999). A major uncertainty in
these studies was the amount of extinction in the UV. Therefore
both extinction and redshift should be included in the SFR
analysis. 
A detailed study of the northern Hubble Deep Field with
\emph{HST} NICMOS, in which both the SFR was determined
individually for galaxies from the $1500 \AA$ UV flux and a
spectral energy distribution template fitting method was applied
to determine redshift as well as extinction, yielded the result
(Thompson et al. 2001) of an increase in SFR between $z= 1 - 2$,
then a fall-off from $z= 2 - 3$ and a plateau region between $z= 3
- 5$ (see Fig.~\ref{fig:sfr}); as a clear result, the SFR was
higher up to redshifts of $\sim 5$ than at present.
Naturally, an enhanced SFR implies an enhanced starburst activity
during the infancy of galaxy evolution and galactic winds are
expected to have been fairly common at that time. Such an
assumption is indeed confirmed by observation. A recent
serendipitous discovery of a galactic wind at $z=5.19$ by Dawson
et al. (2002) revealed a strong asymmetric Ly$\alpha$ emission
line (see Fig.~\ref{fig:gw_la}).
The best fit model line profile resulted in a red- and blueshifted
wing, indicating an outflow velocity of $320 - 360$ km/s. As
will be shown below, such velocities are entirely consistent with
wind speeds derived from modeling the X-ray emission of {\em
local} starburst galaxies. An important consequence of winds is
the pollution of the intergalactic medium (IGM) with chemically
enriched material. In addition to metals, also entropy is added to
the IGM.

This may be an explanation for the so-called ''entropy floor''
postulated by comparing the X-ray luminosity versus temperature
relationship. Going from massive clusters of galaxies
to poor groups there is a deviation from a power law towards the
low mass end. This has been interpreted in terms of an ``entropy
floor'' that dominates the gravitational heating of the poor clusters 
and groups as
the potential wells become shallower (Ponman et al. 1999). Thus
the preheating of the IGM by starburst driven galactic winds
starts to dominate the release of gravitational energy. 
Also, the fairly high
abundances of $Z \sim 0.3 \, Z_\odot$ (e.g., Molendi et al. 1999)
found in the intra-cluster gas may be explained by ejection of
metals from galaxies by massive winds during initial starburst
phases.

In the local universe, evidence for galactic winds mainly stems
from observations of starburst galaxies, such as M\,82 or
NGC\,253. In these more evolved objects, starbursts are thought to
be triggered by a substantial disturbance of the gravitational
potential, e.g.\ by interaction with a companion. Nevertheless, as
it has been pointed out, local starbursts are
similar in origin, evolution, duration and its effects on the IGM
to those at high redshifts (e.g.\ Pettini et al.2000). If one accepts 
these findings as a
working hypothesis, nearby starburst galaxies would serve as ideal
laboratories for studying the effects of starbursts on the early
evolution of galaxies and the interaction with the IGM. Since the
starburst driven outflows are mostly thermal, they are best
observed in soft X-rays.
\section{Observations of local starburst galaxies}
Direct evidence for supernova (SN) heated galactic halos stems
from imaging and spectroscopy of diffuse soft X-ray emission. This
has been impressively demonstrated by a recent XMM-Newton
observation of NGC\,253 (Pietsch et al. 2001).
%
%
%
\begin{figure}[!t]
  \includegraphics[width=\columnwidth,clip]{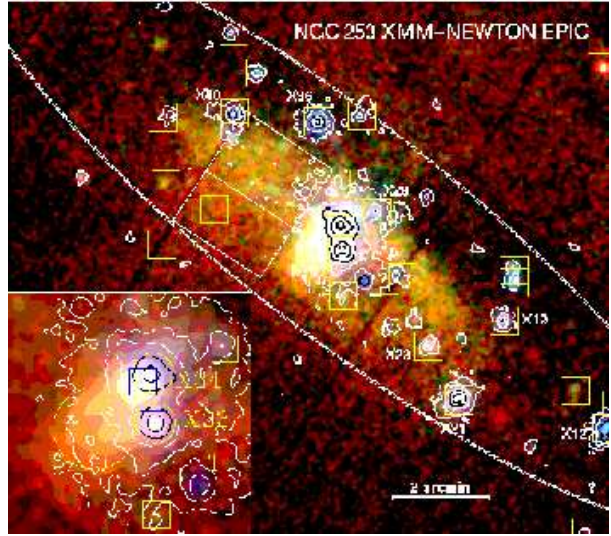}
  \caption{The extended soft X-ray emission of the nearby edge-on
  starburst galaxy NGC\,253, as observed by the EPIC pn and MOS cameras
  onboard XMM-Newton in a three-colour image (red: 0.2 - 0.5 keV, 
  green: 0.5 - 0.9 keV, blue: 0.9 - 2.0 keV; overlayed contours represent the 
  2.0 - 10.0 keV band; Pietsch et al. 2001). For reference the $D_{25}$
  ellipse is plotted to show the optical extension of the 
  gas disk. The inlay is a zoom-in of the nuclear region, with ROSAT detected 
  sources marked by squares.}
  \label{fig:n253_img}
\end{figure}
From Fig.~\ref{fig:n253_img} it can be clearly inferred that (i)
the diffuse emission is soft (cf.\ $0.2 - 0.5$
keV band colour coded in red), and (ii) that the outflow is extended and 
distributed
over the whole disk rather than being constrained solely to the
nuclear region. 
%
\begin{figure}[!t]
  \includegraphics[bb=99 207 473 580,width=\columnwidth,clip]{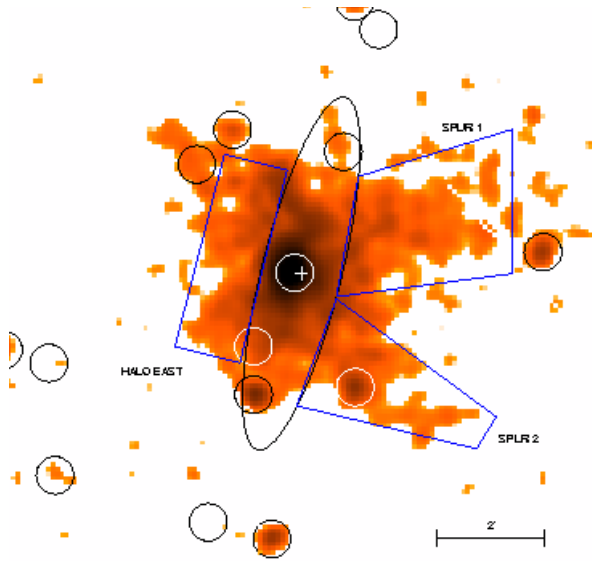}
  \caption{The extended soft X-ray halo of the nearby edge-on
  starburst galaxy NGC3079, as observed by the EPIC pn camera
  onboard XMM-Newton in the 0.2 - 1.0 keV energy band
  (Breitschwerdt et al. 2003). Point sources are shown by
  circles. For reference the $D_{25}$
  ellipse is indicated to show the optical extension of the underlying
  gas disk. }
  \label{fig:n3079_img_soft}
\end{figure}
The nearby (distance 17 kpc) edge-on (inclination to the line of
sight about $85^\circ$) spiral galaxy NGC3079 has been observed
with XMM-Newton for a total of 25 ksec. The galaxy is classified
as an SBc LINER with distinct nuclear activity. The object is an
ideal target for analyzing the diffuse X-ray emission in the disk
and the surrounding galactic halo, since galactic foreground
absorption is low (column density $N_H < 10^{20} {\rm cm}^{-2}$).
The target was chosen to investigate both the problem of starburst
AGN connection, and to see if there was a wide-spread extranuclear
X-ray emission, for which we wanted to derive its spectral
properties. The observation was carried out as part of the XMM SSC
Guaranteed Time Program. Rejection of high background times led to
an overall useable integration time of 16.4 ksec. We detected a
huge soft ($0.2 - 1.0$ keV) X-ray halo
(s.~Fig.~\ref{fig:n3079_img_soft}) extending perpendicular to the
galaxy disk to about 17.5 kpc (Breitschwerdt et al., 2003). 
The emission is therefore expected to be dominated by
thermal lines, which are abundant in the soft band. The spectrum
of a conspicuous emission region indeed proves this point.
Moreover it clearly shows that the galactic starburst must drive a
thermal outflow, since we have strong indications for
collisionally excited oxygen and iron L line complexes in the
spectrum (s.~Fig.~\ref{fig:n3079_spfit}). A further argument for a
halo connected to starforming regions in the underlying gaseous
disk rather than to AGN activity is the softness of the halo,
which completely disappears in the higher energy band (1.0 - 2.0
keV). The morphology of the halo also shows extranuclear spurs
supporting the starburst connection.
%
%
%
\section{Self-consistent dynamical and thermal modelling of outflows}
\begin{figure}[!t]
  \includegraphics[bb=0 0 530 735,angle=-90,width=\columnwidth,clip]{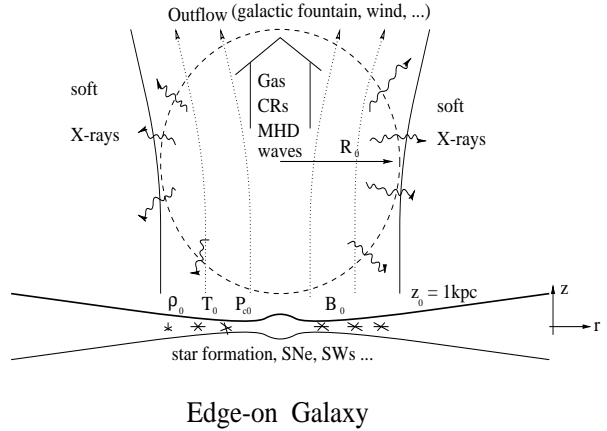}
  \caption{Sketch of the galactic outflow model. The wind is mainly driven by thermal
  pressure, but cosmic rays and MHD waves may also assist and are included self-consistently.}
  \label{fig:eo_hal}
\end{figure}

The injection of SN heated gas into the base of the halo generates
a huge pressure gradient with respect to the ambient IGM.
Consequently, the hot plasma can escape the gravitational
potential well and expand to infinity, provided that the external
pressure is sufficiently low. Otherwise, the expanding hot gas
will become thermally unstable and rain down in the form of
intermediate velocity clouds on the galactic disk (galactic
fountain). This is to be expected if the radiative cooling time
scale, $\tau_{\rm cool} \simeq 3 n \Lambda(T)/(k_B T)$, is much less
than the dynamical flow time scale, $\tau_f(z) = \int_{z0}^{z} d
z^\prime/u(z^\prime)$; in a steady state flow this condition has
to hold at least at the sonic point. Further out the
flow is supersonic and cannot be influenced by the source region;
mass loss is then inevitable.  In a starburst driven flow this is
certainly fulfilled as our calculations show. The model as applied
to an edge-on galaxy is sketched in Fig.~\ref{fig:eo_hal}.
We have tested the hypothesis of a thermally driven superwind by
modelling the outflow with our galactic wind hydrocode, in which
the radiative cooling and X-ray emission is treated
self-consistently with the dynamics in full non-equilibrium (NEI).
In essence this means that we follow the time-dependent ionization
structure of the outflow instead of using the usual assumption of
collisional ionization equilibrium (CIE). The necessity in doing
so has been been extensively discussed in Breitschwerdt \&
Schmutzler (1999). For brevity we just repeat here the main
arguments. Firstly, the assumption of CIE is strictly never
fulfilled. A detailed balancing of the collisional ionization rate
by the recombination rate would require three body collisions (one
electron recombining with an ion, another one carrying away the
excess energy), which are extremely rare in an optically thin and
dilute plasma. Therefore collisional de-excitation is negligible
and the coronal approximation holds. Instead radiative cooling, by
which the excess energy is removed by a photon, is dominating by
far. Except for resonant processes, the absorption probability of
these photons is very low and they escape. But this means that both collisional
ionization and radiative recombination are \emph{cooling
processes} for the plasma, which will hence be driven out of
equilibrium (Shapiro \& Moore, 1976; Schmutzler \& Tscharnuter,
1993). CIE can therefore only hold as an approximation during a
limited period or due to some form of heat input. Secondly, it has
been shown that if the plasma is in a dynamical state (e.g.
expansion), the dynamical time scale can be shorter than any
atomic time scale and the plasma is characterized by a state of
\emph{delayed recombination} (Breitschwerdt \& Schmutzler, 1994). 
Thirdly, there exists a feedback
mechanism between thermal and dynamical evolution. The latter
changes the density, pressure etc. of the flow, which in turn
changes the time-dependent ionization structure and thereby the
radiative energy losses, which crucially determine the dynamics
again. An approximate and efficient way in solving the
hydrodynamics \emph{self-consistently} along with the
non-equilibrium ionization (NEI) structure is to 
calculate the temporal evolution separately and couple them by a
\emph{time-dependent cooling rate} ${\cal L}[n(t),T(t);Z]$,
where $n$, $T$ and $Z$ are the density, temperature and chemical
abundances, respectively (Breitschwerdt \& Schmutzler, 1999). This
iterative procedure converges fairly rapidly. The NEI code treats
the time-dependent atomic processes (collisional ionization,
recombination, autoionization, dielectronic recombination, to name
just a few; for details see Schmutzler \& Tscharnuter, 1993) of
the ten most abundant elements and their respective ionization
stages, and includes at present 1155 lines. The hydrocode has been
described in detail by Breitschwerdt et al. (1991). It entails a
realistic gravitational potential of the galaxy, and the flow is
calculated in steady-state in a flux tube geometry, which is a
fair representation of the transition from plane parallel flow
close to the disk to spherical divergence far away. We expect that
steady-state is justified \emph{during} the starburst and as long
as there are no dramatic changes at the inner boundary conditions.
Fully time-dependent simulations, using the STARBURST99 code, are
underway (Dorfi \& Breitschwerdt 2003), and confirm this by and
large, but also show that abrupt changes in the SN rate lead to
propagation of shocks, reheating of the halo and particle
acceleration out there to energies well in excess of $10^{15}$ eV
reached in individual SNRs. This has to be borne in mind when interpreting
the spectra.

\begin{figure}[!t]
  \includegraphics[bb= 0 0 585 390,width=\columnwidth, clip]{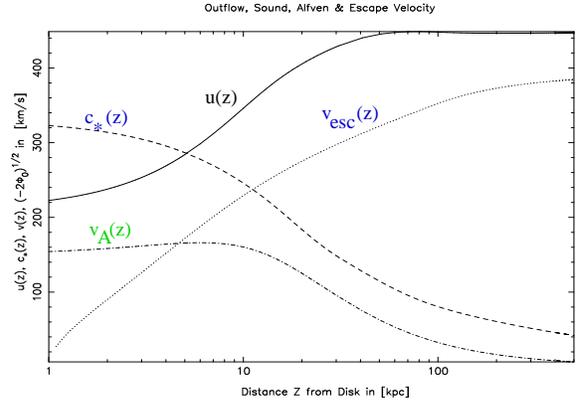}
  \caption{Dynamically and thermally self-consistent outflow model of the starburst
  driven galactic wind of NGC\,3079 (Breitschwerdt et al. 2003). The solid line
  is the outflow velocity $u(z)$, $v_A(z)$ is the Alfv\'{e}n velocity (dash-dotted line),
  $c_*(z)$ is the generalized speed of sound (dashed line) and $v_{\rm esc}(z)$
  (dotted line) is the escape speed from the gravitational potential of the galaxy. }
  \label{fig:n3079_outflow}
\end{figure}

\begin{figure}[!t]
  \includegraphics[bb= 65 35 560 710,angle=-90,width=\columnwidth, clip]{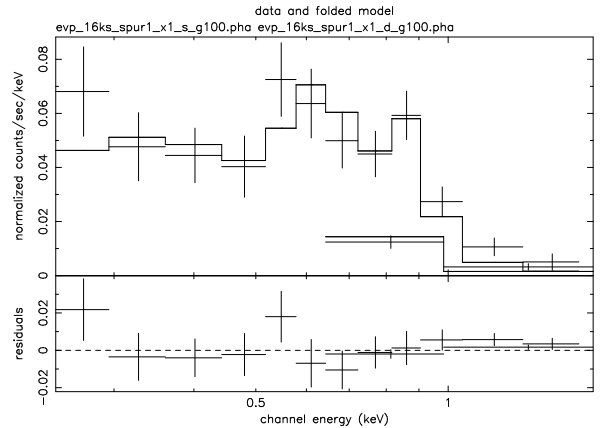}
  \caption{Fit ($\chi^2 = 1.2$) of the Epic pn observed spectrum (crosses =
  error bars) of the soft X-ray halo emission of NGC3079 between $0.2 - 2$
  keV obtained with a self-consistent non-equilibrium emission outflow
  model (solid line; Breitschwerdt et al. 2003). }
  \label{fig:n3079_spfit}
\end{figure}

In order to compare the simulated spectra directly with the
observations, we have folded them through the EPIC pn instrumental
response. Taking the instrumental field of view of $\sim 30^\prime
\times 30^\prime$, we have projected the outflow cone at the
galaxy's distance (see Fig.~\ref{fig:eo_hal}) onto it. Thus we do
not only reproduce the form of the spectrum, but also the absolute
count rate. We even went one step further,
and instead of comparing the observed and synthetic spectra by
eye, we have used the latter as input for XSPEC to actually
\emph{fit} the observed spectrum. The result shows
(s.~Fig.~\ref{fig:n3079_outflow}) that the halo indeed exhibits a
''multi-temperature structure'', fully consistent with an NEI
outflow, in which supernova heated gas is injected into the halo
at temperatures of $3.6 \times 10^6$ K and density $5 \times
10^{-3} \, {\rm cm}^{-3}$ at an initial velocity of about 220 km/s
(see Fig.~\ref{fig:n3079_outflow})
to give a satisfactory fit with $\chi^2_{red} \approx 1.2$. A
closer inspection of Fig.~\ref{fig:n3079_spfit} reveals two distinct
humps at $0.6$ keV and $0.9$ keV, caused by the \ion{O}{7}, \ion{O}{8} and
Fe-L line complexes, respectively.
We emphasize that a single temperature CIE model, as e.g. by
Raymond \& Smith (1977) or a so-called MEKAL model are not able to
reproduce the spectrum in a statistically acceptable way. The
reason is, because CIE models are extremely temperature sensitive,
each hump in our spectrum would require a different CIE
temperature. In an attempt to achieve a fit, although not a very
good one, a common error is to decrease the element abundances to
extremely low values (e.g. a few percent solar, as been advocated
in some publications) in order to get rid of prominent lines which
would enforce a specific excitation temperature. We consider this
as a pure \emph{artefact} of the fit procedure, and moreover as
\emph{unphysical}, because an outflow that is mass loaded by
chemically enriched material from the starburst region should have
rather high than low metallicities.
\section{Conclusions}
Star formation is a highly nonlinear process, and it has been
known for some time, that for example it can be triggered by
nearby SN explosions. On the other hand the rate cannot increase
indefinitely, because the exhaustion of gas fuel renders it more
and more inefficient, thereby forcing it into a self-regulating
cycle. The highest rates known to date are realized by
galaxy-galaxy interactions (with mergers as an extreme form), and
starbursts as a common manifestation. In particular in the early
universe when the average distance between galaxies was much
smaller, starbursts could dominate their integrated luminosity.
Both local and distant starburst galaxies show that the enhanced
star formation rate is accompanied by a thermally driven outflow,
which can be traced to large distances in the soft X-rays.
Although only a few percent of the energy released is radiated in
this wavelength range, the presence of lines of highly ionized an
abundant species such as oxygen and iron in the spectrum serves as a
clear fingerprint for the generation of a hot plasma and its 
dynamical evolution as a galactic wind. Therefore spectral
modelling of outflows provides important quantitative information
of mass loss rates, chemical enrichment of the IGM, and, possibly
a contribution to the WHIM (warm hot intergalactic medium) that
has been invoked to explain the so-called missing baryon problem
(Cen \& Ostriker 1999). Last but not least, the effect of a
starburst induced superwind on the further evolution of the host
galaxy should not be underrated. Venting hot material away removes
a lot of entropy thus ensuring a continuation of star formation in
a more quiescent fashion in the disk.

\paragraph{Acknowledgement.} It is an honour and a pleasure to thank John 
Dyson, who has been a mentor during my stay as a post-doc at 
the Department of Astronomy in Manchester, and ever since a friend, 
whose advice and kindness I have always appreciated. I thank my colleagues 
Drs. W. Pietsch and A. Vogler for permission to reproduce unpublished 
figures.
I am grateful to Jane Arthur and 
the Organizing Committee for their invitation to an excellent meeting and 
for financial support.
I also thank the Max-Planck-Institut f\"ur extraterrestrische 
Physik for financial support and the Deutsche Forschungsgemeinschaft for 
a travel grant. The XMM-Newton project is supported by the Bundesministerium 
f\"ur Bildung und Forschung/Deutsches Zentrum f\"ur Luft- und Raumfahrt
(BMBF/DLR), the Max-Planck-Gesellschaft and the Heidenhain-Stiftung.


\begin{thebibliography}

\bibitem{} Breitschwerdt D., McKenzie J.F., V\"olk H.J., 1991, A\&A, 245, 79

\bibitem{} Breitschwerdt D., Pietsch, W., Vogler, A., Read, A.M, 
Trinchieri, G., 2003 (in preparation)

\bibitem{} Breitschwerdt, D., Schmutzler, T., 1994, Nature 371, 774

\bibitem{} Breitschwerdt, D.,  Schmutzler, T., 1999, A\&A 347, 650

\bibitem{} Cen, R., Ostriker, J.P., 1999, ApJ 514, 1

\bibitem{} Chevalier, R.A, Clegg, A.W., 1985, Nature 317, 44

\bibitem{} Dawson, S., Spinrad, H., Stern, D., Dey, A., van
Breugel, W., de Vries, W., Reuland, M. 2002, ApJ 570, 92

\bibitem{} Dorfi, E.A., Breitschwerdt, D., 2003 (in preparation)

\bibitem{} Heckman, T. 1997, in: "Cosmic Origins of Galaxies,
Planets, and Life", eds. J.M. Shull, C. Woodward, H. Thronson, ASP


\bibitem{} Kulsrud, R.M., Pearce, W.D., 1969, ApJ 156, 445


\bibitem{} Madau, P., Ferguson, H.C., Dickinson, M.E., Giavalisco,
M., Steidel, C.C., Fruchter, A., 1996, MNRAS 283, 1388

\bibitem{} Madau, P., Pozzetti, L., Dickinson, M., 1998, ApJ 498,
106

\bibitem{} Molendi, S., de Grandi, S., Fusco-Femiano, R., et al.,
1999, ApJ 525, L73

\bibitem{} Pettini, M., Steidel, C.C., Adelberger, K.L., Dickinson, M., 
Giavalisco, M., 2000, ApJ 528, 96

\bibitem{} Pietsch, W., Roberts, T.P., Sako, M., et al., 2001,
A\&A 365, L174

\bibitem{} Ponman, T.J., Cannon, D.B., Navarro, J.F., 1999, Nature
397, 135

\bibitem{} Raymond, J.C., Smith, B.W. 1977, ApJS 35, 419

\bibitem{} Searle, L., Sargent, W.L.W., Bagnuolo, W.G., 1973, ApJ
179, 427

\bibitem{} Schmutzler, T., Tscharnuter, W.M. 1993, A\&A 273, 318

\bibitem{} Shapiro, P.R., Moore, R.T., 1976, ApJ 207, 460

\bibitem{} Steidel, C.C., Adelberger, K.L., Giavalisco, M.,
Dickinson, M., Pettini, M., 1999, ApJ 519, 1


\bibitem{} Thompson, R.I., Weymann, R.J., Storrie-Lombardi, L.J.
2001, ApJ 546, 694

\bibitem{} V\"olk, H.J., Aharonian, F.A., Breitschwerdt, D., 1996,
Sp. Sci. Rev. 75, 279

\end{thebibliography}
\end{document}